\documentclass[aps,preprintnumbers,nofootinbib,superscriptaddress]{revtex4}

\usepackage{graphicx}
\DeclareGraphicsExtensions{.pdf}

\usepackage{amsfonts,amssymb,amsmath}
\usepackage{amsthm}

\newcommand{\Prob}{{\rm Prob}}
\newcommand{\comment}[1]{}

\theoremstyle{plain}

\theoremstyle{definition}

\begin{document}

\title{Quantum Tasks in Minkowski Space}

\author{Adrian \surname{Kent}}
\affiliation{Centre for Quantum Information and Foundations, DAMTP, Centre for
  Mathematical Sciences, University of Cambridge, Wilberforce Road,
  Cambridge, CB3 0WA, U.K.}
\affiliation{Perimeter Institute for Theoretical Physics, 31 Caroline Street North, Waterloo, ON N2L 2Y5, Canada.}

\date{ April 2012}

\begin{abstract}

The fundamental properties 
of quantum information and its applications to computing and
cryptography have been greatly illuminated by considering 
information-theoretic tasks that are provably possible or impossible
within non-relativistic quantum mechanics.    
I describe here a general framework for defining 
tasks within (special) relativistic quantum theory and illustrate it with 
examples from relativistic quantum
cryptography and relativistic 
distributed quantum computation.  
The framework gives a unified description of all tasks previously 
considered and also defines a large class of new questions about the properties
of quantum information in relation to Minkowski causality.  
It offers a way of exploring interesting new fundamental tasks and
applications, and  
also highlights the scope for a more systematic understanding of
the fundamental information-theoretic properties of relativistic
quantum theory.   
\end{abstract}

\maketitle

\section{Introduction}
\label{sec:intro}

\subsection{Theoretical motivations}

Although some of the fundamental properties of quantum theory -- for
example, the superposition principle -- were understood very early,
other key insights were made only later.  Quantum entanglement was
first described by Schr\"odinger \cite{entanglement} only in the
1930s; it was not till the 1960s that Bell showed that quantum theory
violates local causality
\cite{belllocbeables,bellloccaus,bellbook,CHSH}; some other important
aspects of the delicate relationship between the general quantum
measurement postulates and the no-signalling principle were not
completely understood until even more
recently \cite{gisinone,gisintwo,czachor,aknlwithouts,infocaus}.  These
and other features of quantum theory have been greatly illuminated in
recent decades by the development of quantum computing, quantum
cryptography and quantum information theory, which have inspired a
perspective on quantum theory in terms of tasks and resources
involving physical information.

Conversely, considering the possibility and impossibility 
of various quantum tasks
led to significant discoveries in quantum communication (e.g.
\cite{teleportation}) and quantum cryptography (e.g.
\cite{wz,dieks,mayersprl,lochauprl}).  
Although mathematically trivial, the 
quantum no-cloning theorem \cite{wz,dieks}, 
nonetheless encapsulated a
fundamental fact about quantum theory.  It inspired several other
significant results, including independent proofs of the impossibility
of determining an unknown quantum state \cite{day} and the
impossibility of distinguishing between non-orthogonal
states \cite{yuen}, the no-deletion theorem \cite{braunsteinpati}, the
no-broadcasting theorem for mixed states \cite{nobroadcast}, a general
no-cloning theorem incorporating several of these
results \cite{lindblad} and a proof that it is impossible to clone with
partial ancillary information \cite{jozsa}.  A further significant
extension was the introduction of the idea of partial fidelity
cloning \cite{bh}, and the discovery of universal algorithms for
attaining the best possible state-independent fidelities for $M
\rightarrow N$ partial cloning
 \cite{gisinmassar,bem,werner,keylwerner}.  Work on information
causality \cite{infocaus} has given further insight into quantum theory
and its relationship with special relativity.

All of these results shed light on the relationship
between quantum theory and special relativity.  Several of them are crucial
to our current understanding.   However, they all describe features already
evident in non-relativistic quantum mechanics.
As we currently understand things, relativistic quantum 
theory is closer to the true description of nature than quantum
mechanics.  So, there remains a compelling motivation to identify
properties and principles that are intrinsic to relativistic quantum
theory.   This is especially true since 
we understand relativistic quantum theory so poorly compared to
quantum mechanics.  We have an informal intuitive understanding of
many features of Lorentz invariant quantum field theories with local
interactions, but as yet no rigorous definition of any non-trivial
relativistic quantum theory.  One might hope to make
progress here by identifying the principles that such a theory must
satisfy. 

The no-summoning theorem \cite{nosummoning} represents a step in
this direction.  To define the relevant task, we need to consider two
agencies, Alice and Bob, with representatives distributed throughout
space-time.  Bob prepares a localized physical state, whose identity
is known to him but kept secret from Alice, and hands it over to her
at some space-time point $P$.  At some point $Q$ in the causal future
of $P$, he {\it summons} the state -- i.e. he asks Alice to return it.
The location $Q$ may be known to Bob from the start, but is kept
secret from Alice until the request is made.  It is easy to show that,
if this task is modelled within relativistic quantum theory -- i.e.
quantum theory in Minkowski space -- then in general she will not be
able to comply.  Interestingly, however, she {\it can} comply if the
underlying theory is taken to be either relativistic classical theory
or non-relativistic quantum theory \cite{nosummoning}.  In this sense,
the no-summoning theorem identifies an information-theoretic principle
that we believe holds true in relativistic quantum theory and in
nature but that does not hold true in
either of the theories that relativistic quantum theory replaces
(or would replace if rigorously defined).

It seems natural, then, to try to find other tasks that teach us
more about relativistic quantum theory.   If a sufficiently 
complete list can be found, this might even be 
a strategy for rigorously defining relativistic quantum theory, 
as precisely that theory that allows one list of tasks 
(the ``possible'' tasks) and precludes a second list
(the ``impossible'' ones).   It seems natural too to try to
find a general framework that not only includes all the familiar
possible and impossible information-theoretic tasks that
characterize non-relativistic quantum theory, but also includes 
the task of summoning and (presumably) many others that
characterize relativistic quantum theory.   This paper
proposes such a framework. 

\subsection{Cryptographic motivations}

Quantum theory and the relativistic no-signalling principle both give
ways of controlling information, in the sense that someone who creates
quantum information somewhere in space-time can rely on strict limits
both on how much information another party can extract and on where
they can obtain it.  While standard quantum cryptography (e.g.
\cite{wiesner,BBeightyfour,ekert,qsecrets,anonqcomm,qmultiparty}) uses
only the properties of quantum information, an increasingly long list
of applications illustrate the added cryptographic power of the
relativistic no-signalling principle, either alone (e.g.
\cite{power,kentrel,kentrelfinite,colbeckkent}), or when combined with
quantum information (e.g.
\cite{taggingpatent,malaney,kms,chandranetal,buhrmanetal,kenttaggingcrypto,beigikoenig,BHK,BKP,AGM,
  AMP,ABGMPS,PABGMS,McKague,MRC,Masanes,HRW2,MPA,HR,ColbeckThesis,PAMBMMOHLMM,ckrandom,DIBC,BCK,
  nosummoning,bcsummoning,otsummoning,bcmeasurement,kwiat}).

A new relativistic quantum cryptographic technique was 
recently introduced, inspired by the no-summoning theorem
\cite{nosummoning}, in which one agency (Alice or $A$) sends a quantum state,
supplied by and known to
another agency (Bob or $B$) but unknown to $A$, at light speed $c$ in one of 
a number of possible directions.  The term `` agency'' here is used to 
emphasize that Alice and Bob are not single isolated individuals: they
have representatives 
distributed at various points in space-time.
We assume all these representatives are loyal and
act according to the instructions of their agency; however, Alice and
Bob do not trust one another.    
The task is securely implemented if $A$'s chosen direction is concealed
from $B$ until $A$ chooses to return the state.   

This technique
gives, inter alia, a provably unconditionally secure protocol for 
the cryptographic primitive of bit commitment
\cite{bcsummoning} and a way of transferring data at a location unknown
to the transferrer \cite{otsummoning}.
Other techniques for secure bit commitment using relativistic
signalling constraints alone \cite{kentrel,kentrelfinite} or
combined with the properties of quantum information
\cite{bcmeasurement} have also been developed.  

Another class of applications of quantum information in Minkowski
space that has recently attracted much attention involves schemes
for identifying, verifying and/or exploiting cryptographically
the position of a distant object. 
Perhaps most fundamental task in this class is quantum tagging,
also callled quantum position authentication, which involves using 
communications from distant sites to verify the object's 
location.  An unconditionally secure scheme for quantum  
tagging was recently proposed  \cite{kenttaggingcrypto}, 
following earlier proposals for conditionally secure quantum tagging 
schemes \cite{taggingpatent,malaney,buhrmanetal,kms}
based on slightly weaker security
assumptions.   A large class of schemes for more 
general tasks in position-based quantum cryptography
\cite{buhrmanetal} have also been proposed.   

These cryptographic applications give further motivations for 
defining an abstract framework for information-theoretic tasks in relativistic
quantum theory.  

First, it seems very likely that there are many more
interesting relativistic quantum cryptographic applications to be
discovered, and a more systematic way of defining and classifying 
quantum information-theoretic tasks in Minkowski space seems likely
to be helpful in finding them.   

Second, it is already clear from
the existing applications that we really need a rigorous general
way of {\it defining} quantum cryptographic tasks in Minkowski space.

For example, an apparently slight difference in the definition of 
the task of quantum tagging \cite{kms,kenttaggingcrypto} 
translates into a cryptographically
relevant difference in the security assumptions, with the 
consequence that unconditionally secure quantum tagging is
provably possible in one security scenario \cite{kenttaggingcrypto}
and provably not in another \cite{buhrmanetal}.  

New subtleties also arise in the definition of bit commitment
in Minkowski space.  We use these points below to illustrate 
the framework and its uses.  

\subsection{Quantum computational motivations} 

Quantum computations take place over distributed
networks, which may accept inputs of classical or quantum data 
from sources outside the network.   Such networks (for example
for stock and other market trading) will presumably ultimately
be large scale, extending over the Earth and beyond, and the
signalling constraints implied by Minkowski causality will 
be computationally relevant.   Toy examples show that 
significant efficiency gains can be made by using teleportation,
secret sharing and other applications of quantum information
processing.   However, we lack a 
theory of efficient quantum computational network design 
in Minkowski space that allows us to generate optimal or near-optimal 
networks for any given task, or to prove that a given network
is (nearly) optimal for a given task.  The framework we set 
out allows such questions to be defined and explored.  

\section{Quantum tasks in Minkowski space}

We define tasks for a single agency, Alice, who may, unless otherwise
stipulated, have agents distributed
throughout spacetime.  The agents are, unless otherwise stipulated,
able to send classical and quantum signals to one another along
any lightlike or timelike line in Minkowski space.   
For the moment we suppose that no restrictions are stipulated.

The tasks presuppose {\it oracles} distributed in Minkowski
space that supply finite quantities of classical or quantum
information at a finite
set of points $\{ P_1 , \ldots , P_m \}$. 
We denote the information supplied at $P_j$ by $I_j$, which may be 
either a finite classical signal -- without loss of generality an
integer in the range $\{ 1 , \ldots , d_j \}$ -- or a quantum state $\rho$ in a
$d_j$-dimensional Hilbert space.   The input quantum states may be
entangled with one another and/or with some other systems not
accessible
to Alice.  We  
take $d_j$ to be finite (in either case) unless otherwise stated. 

The points $P_j$ need not all be distinct; a classical and quantum
signal can be supplied at the same point.  The label $j$ is used
for our notational convenience in defining the task
but is not (necessarily) supplied to Alice:
she simply receives some information $I_j$ at some point $P_j$,
without any indication that $P_j$ or $I_j$ are the $j$-th elements
of the relevant sets.  
The values $\{ P_1 , \ldots , P_m \}$ and $\{ I_1 , \ldots , I_m \}$
are the task {\it inputs}. 

Alice's does not generally know in advance the value 
of $m$, the identity of the
points  $\{ P_1 , \ldots , P_m \}$, the classical or quantum 
nature of the signals, or the numbers $d_j$.  However, she does know 
the probability distribution from which these values are all drawn.    
So, from Alice's perspective, she is supplied with random data at 
random points in space-time.  Any agent outside the future light cone
of one of the chosen points $P_j$ will thus generally have 
only probabilistic information
about the likelihood of information being supplied at any point in the
neighbourhood of $P_j$, and about the information, if supplied, taking
any given form $I_j$. 

A protocol, given in advance to Alice (i.e. to all of her agents who may
potentially be involved in the task), determines the {\it
  outputs} as functions of the inputs.
The outputs take the form of some finite list $\{ Q_1 , \ldots , Q_n
\}$ of space-time points, together with classical or quantum
information $\{ J_1 , \ldots , J_n \}$ -- integers in the range 
$\{ 1 , \ldots , e_j \}$ or a state in an $e_j$-dimensional Hilbert
space -- that she is supposed to produce
at the corresponding points.  The required outputs may be entangled.
Alice does not generally know before
the task begins 
the value of $n$, the identity of the points $\{ Q_1 , \ldots , Q_n
\}$, the classical or quantum nature of the output signals required
at these points, or the numbers $e_j$.   These are all deducible 
once all the inputs have been received and collated.  However, even
if the task can be completed, it may not necessarily be possible to
complete it by propagating all the
inputs to some point $X$, calculating all the outputs at $X$, and then
sending signals to the $Q_j$ to produce the outputs there: this depends
on the space-time geometry.   
The output labels $j$, like the input labels, are only for notational
convenience.  So long as she produces the required information at each output
point, Alice completes the task: she 
does not also need to identify the location of each output point in an
ordered list.

\begin{figure}[hb]
\centering
\includegraphics[scale=0.4]{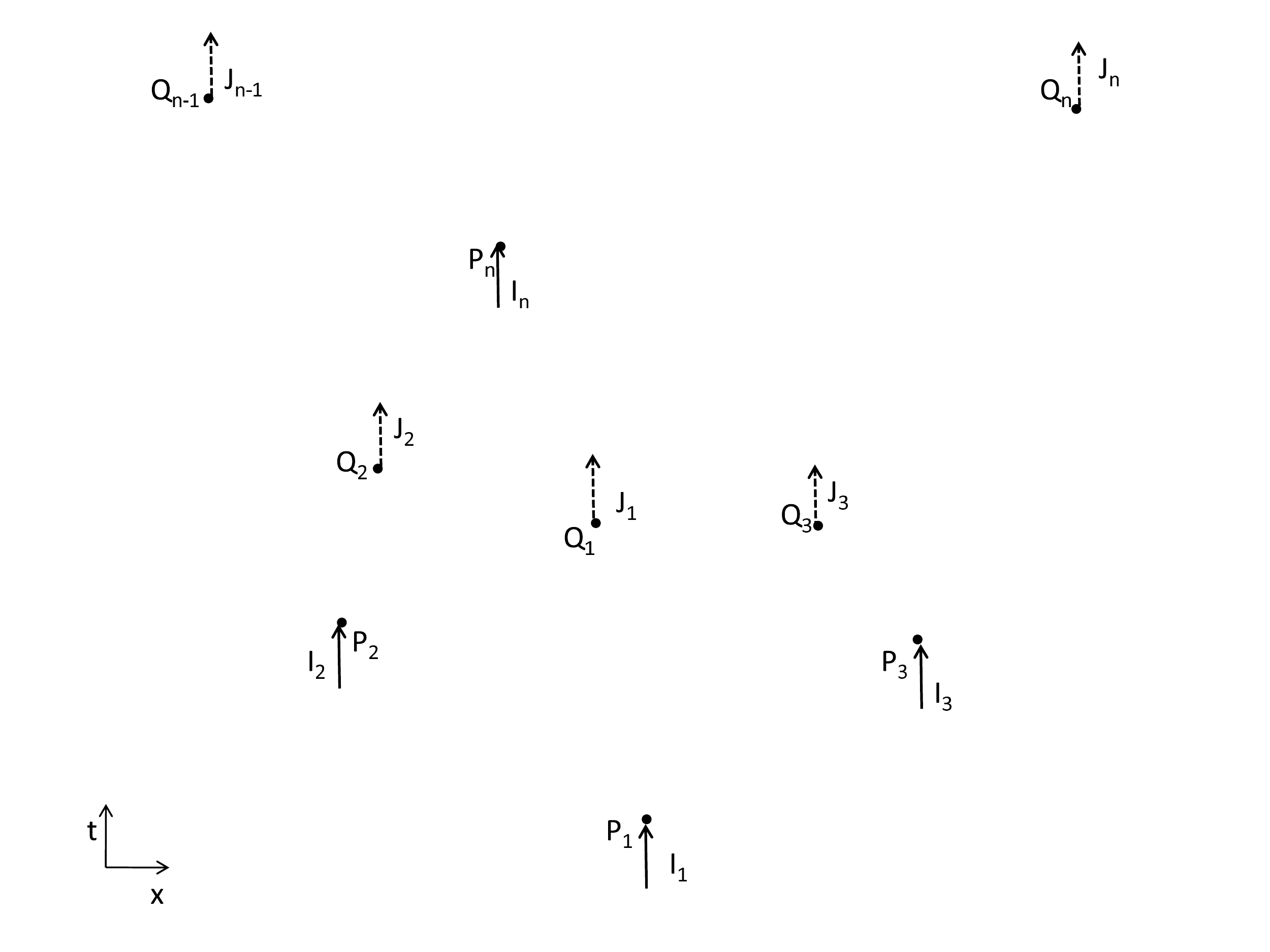}
\caption{An illustration of a relativistic quantum task
in $1+1$ dimensions with no restrictions on the location 
of Alice's agents or their signalling, beyond those 
implied by Minkowski causality.  
Alice receives inputs $I_1 , \ldots , I_m$ at points $P_1 , \ldots , P_m$. 
Following a prearranged protocol, she is required to 
calculate output points $Q_1 , \ldots , Q_n$ and produce the
output data $J_1 , \ldots , J_n$ there.} 
\end{figure}

\newpage

\section{Tasks with no restrictions on Alice}

\subsection{Fundamental principles}

The class of relativistic quantum tasks described by this
framework includes some familiar examples whose (im)possibility is
well understood. 

The simplest illustration is the (im)possibility of signalling, 
depending whether or not the required signal would be superluminal.
To represent this in our framework, suppose 
Alice receives a classical or quantum
input $I_1$ at $P_1$, drawn from a non-trivial probability
distribution,
and is required to produce the same
information $J_1 = I_1$ at a point $Q_1$, where $P_1$ and
$Q_1$ are both determined by the protocol and so known in
advance to Alice.   Minkowski causality implies that this
is possible if and only if $Q_1$ is lightlike or timelike
separated from $P_1$.   

\begin{figure}[h]
\centering
\includegraphics[scale=0.4]{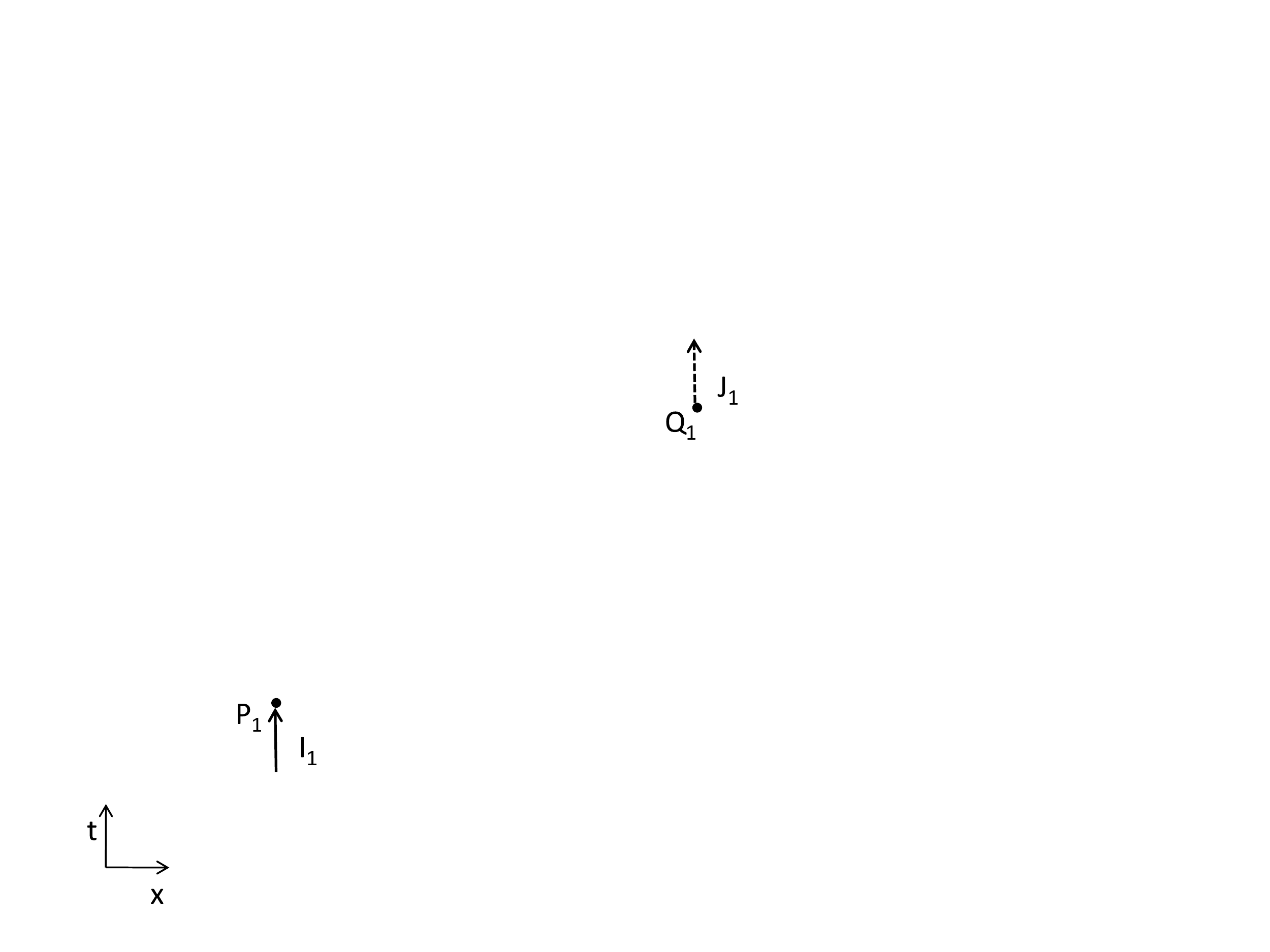}
\caption{The impossibility of superluminal signalling
represented in our framework. 
Alice receives input  $I_1$ at point $P_1$ and
is required to output the same information, $J_1 = I_1$
at point $Q_1$.  She can comply only if $Q_1$ belongs to
the future light cone of $P_1$.}
\end{figure}

\newpage

We can also represent a Bell experiment in a familiar way in this framework,
as a two-input two-output task.   Alice (now represented
by two spacelike separated agents) receives input bits
$I_1$ and $I_2$ at points $P_1$ and $P_2$. She is required
to generate output bits $J_1$ and $J_2$ at points $Q_i$ in
the near future of the respective $P_i$, in such a way
that
$$
\Prob ( J_1 \oplus J_2 = I_1 \cdot I_2 ) > \frac{3}{4} \, .
$$
She can comply provided 
that her agents share entanglement, but not otherwise
\cite{CHSH,hardy}. 

\begin{figure}[h]
\centering
\includegraphics[scale=0.4]{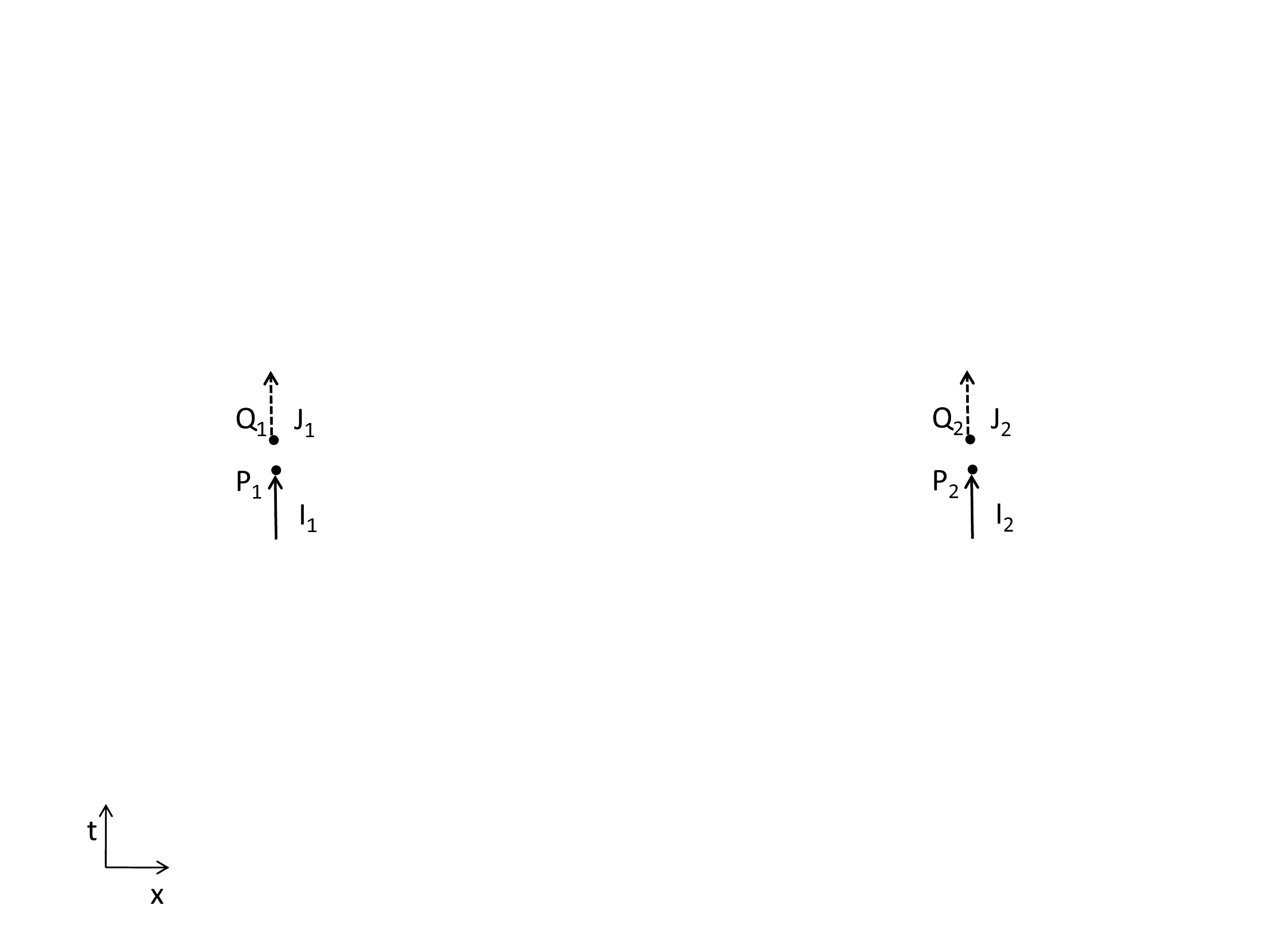}
\caption{A Bell experiment
represented in our framework. 
Alice receives input bits  $I_1$ and $I_2$ at spacelike points $P_1$ and
$P_2$.  She is required to output bits $J_1$ and $J_2$ at points
$Q_1$ and $Q_2$ close to (and timelike separated from) $P_1$ and $P_2$
respectively, in such a way that $ \Prob ( J_1 \oplus J_2 = I_1 \cdot
I_2 ) > \frac{3}{4} \, $. 
  She can comply only if she has agents in the vicinity of $P_1$ and
  $P_2$ that share entanglement.  }
\end{figure}

\newpage

The no-cloning theorem also has a simple representation: 

\begin{figure}[h]
\centering
\includegraphics[scale=0.4]{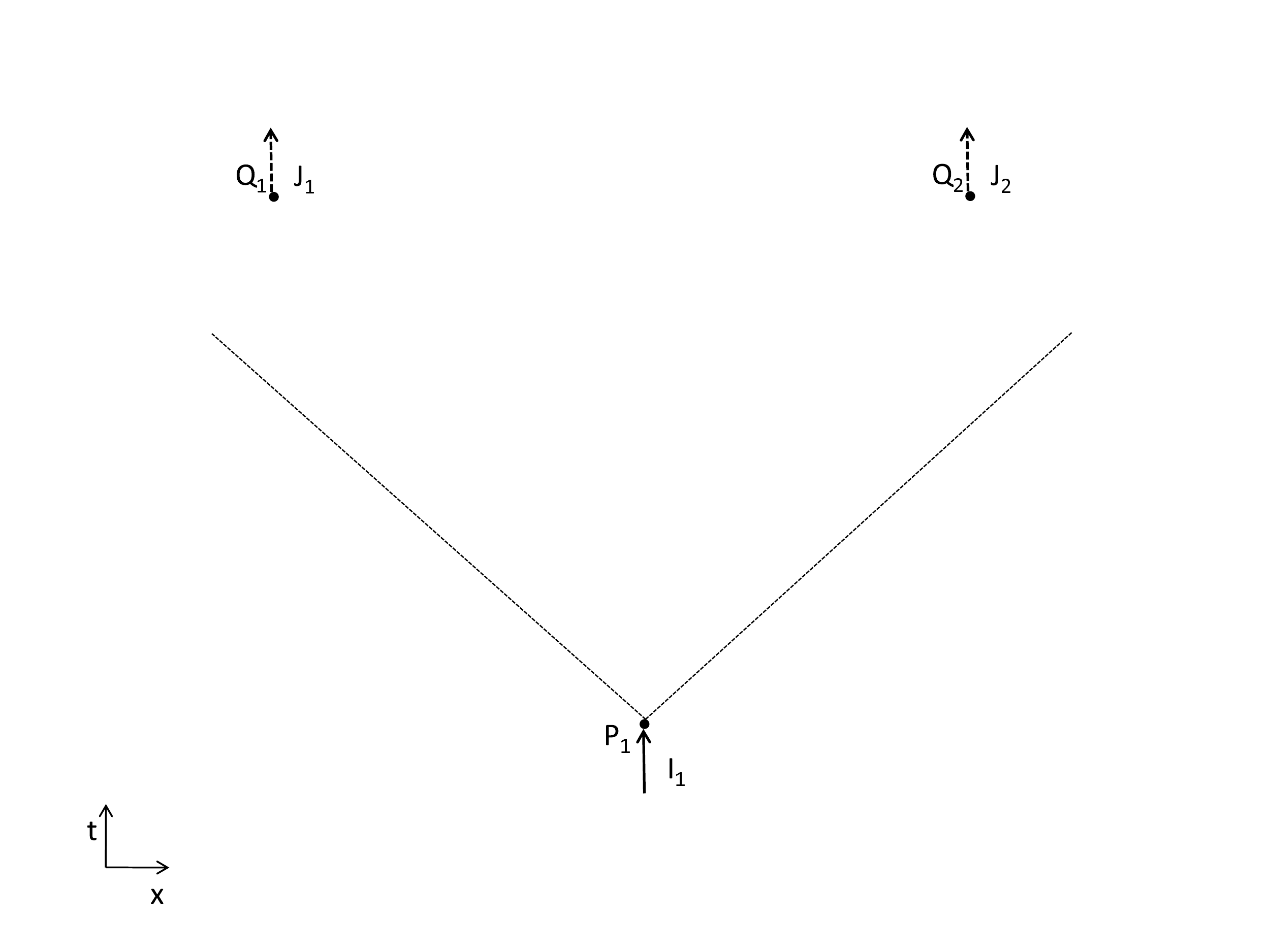}
\caption{The no-cloning theorem 
represented in our framework. 
Alice receives the input quantum state $I_1$ at point $P_1$.  
She is required to output two copies $J_1 = I_1$ and $J_2 = I_1$ at
prestipulated points 
$Q_1$ and $Q_2$ that are timelike separated from $P_1$ and spacelike
separated from one another. 
She cannot comply.   
}\end{figure}

\newpage

One form of the no-summoning theorem \cite{nosummoning} is represented
as follows.   Alice receives input $I_1$, a qudit whose state is
unknown to her, at point $P_1$, which we take to be the origin in
Minkowski space.    At some point $P_2$, whose time
coordinate is $t- \delta$, where $0 < \delta \ll t$. she receives a 
further input, which equals the space coordinate $x_2$ of $P_2$. 
She is required to return the qudit as her output
$J_1$ at the point $Q_1 = (x_2 , t)$. 
She knows in advance the probability distribution for $P_2$ -- 
for example it may be given by the uniform distribution on all 
coordinates $x_2$ satisfying $ | x_2 | \leq t - \delta$, i.e. 
all coordinates corresponding to points lying in the causal
future of $P_1$ at time $t - \delta$.  However, she does not
know which $P_2$ is drawn from this distribution in advance. 

\begin{figure}[h]
\centering
\includegraphics[scale=0.4]{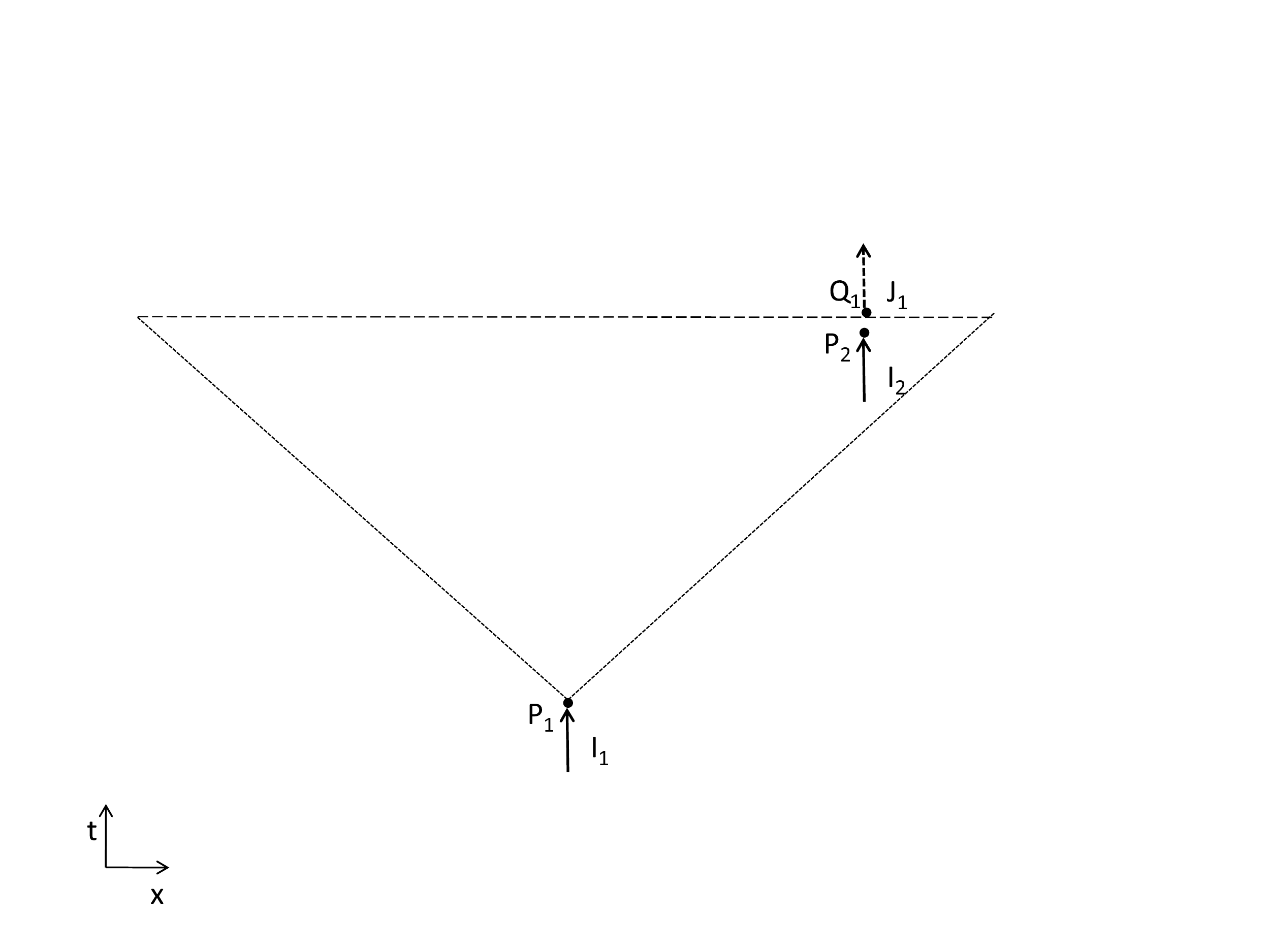}
\caption{A version of the no-summoning theorem 
represented in our framework. 
Alice receives the input quantum state $I_1$ at point $P_1$.  
She is required to return as output the state at some randomly chosen
point $Q_2$, which is identified to her shortly in advance by
an input at $P_2$.  She cannot generally comply.   
}\end{figure}

\newpage

\subsection{Cryptographic tasks}

Relativistic quantum cryptography allows interesting tasks to be
defined (e.g. \cite{otsummoning}) that make no real sense in a 
non-relativistic model.   For example, the 
fact that information is guaranteed to be released at some 
particular point in space has no cryptographic significance
if there is no upper limit on signal speed, since the information
can instantaneously be shared everywhere.  In contrast, a 
guarantee that information is released at some point in space-time
guarantees genuine constraints on its dissemination.  

Considering cryptography in Minkowski space also requires some
reappraisal and refinement of the definitions
of familiar non-relativistic tasks.  We focus here
on bit commitment, which illustrates the point well.
Roughly speaking, in 
non-relativistic classical bit commitment, one party (Alice) {\it commits}
herself to a bit value $b$ at some given time $t$, and then may
choose to {\it unveil} the bit value to the other (Bob) at a later
time $t'$.  In an ideal protocol, the unveiling guarantees to Bob
that Alice was genuinely committed from time $t$ onwards. 
She should have no strategy that allows her to decide on 
the value of $b$ at any time $T>t$ and still produce a valid
unveiling of $b$ at time $t'$.  

Evidently, to extend this definition to Minkowski space we need to
refer to space-time points rather than time coordinates.  It also
turns out to be very useful to consider protocols
\cite{bcsummoning,bcmeasurement} at which the unveiling takes place
not at a single point in space-time but at a {\it set} of space-like
separated points.

But there is another issue.  To define properly what we mean by secure
bit commitment in Minkowski space, we also need
to consider the possibility that Alice's {\it commitment choice} could be
made not at a unique point but through the coordinated actions of 
her agents at a set of points.  In
realistic applications one would normally restrict attention to
the actions of finite sets of agents, and so to assume this set
is finite.     It is theoretically useful, though, also to allow
the set to be infinite, for instance some region 
on a space-like hypersurface. 

To see why coordinated commitment strategies by separated
agents pose new security issues, consider first the bit commitment protocols
of Refs. \cite{bcsummoning,bcmeasurement}.   We can represent the
information flow in these protocols within our framework, by
letting Bob (the recipient) play the role of the oracle, as in the
diagram below.  

\begin{figure}[h]
\centering
\includegraphics[scale=0.4]{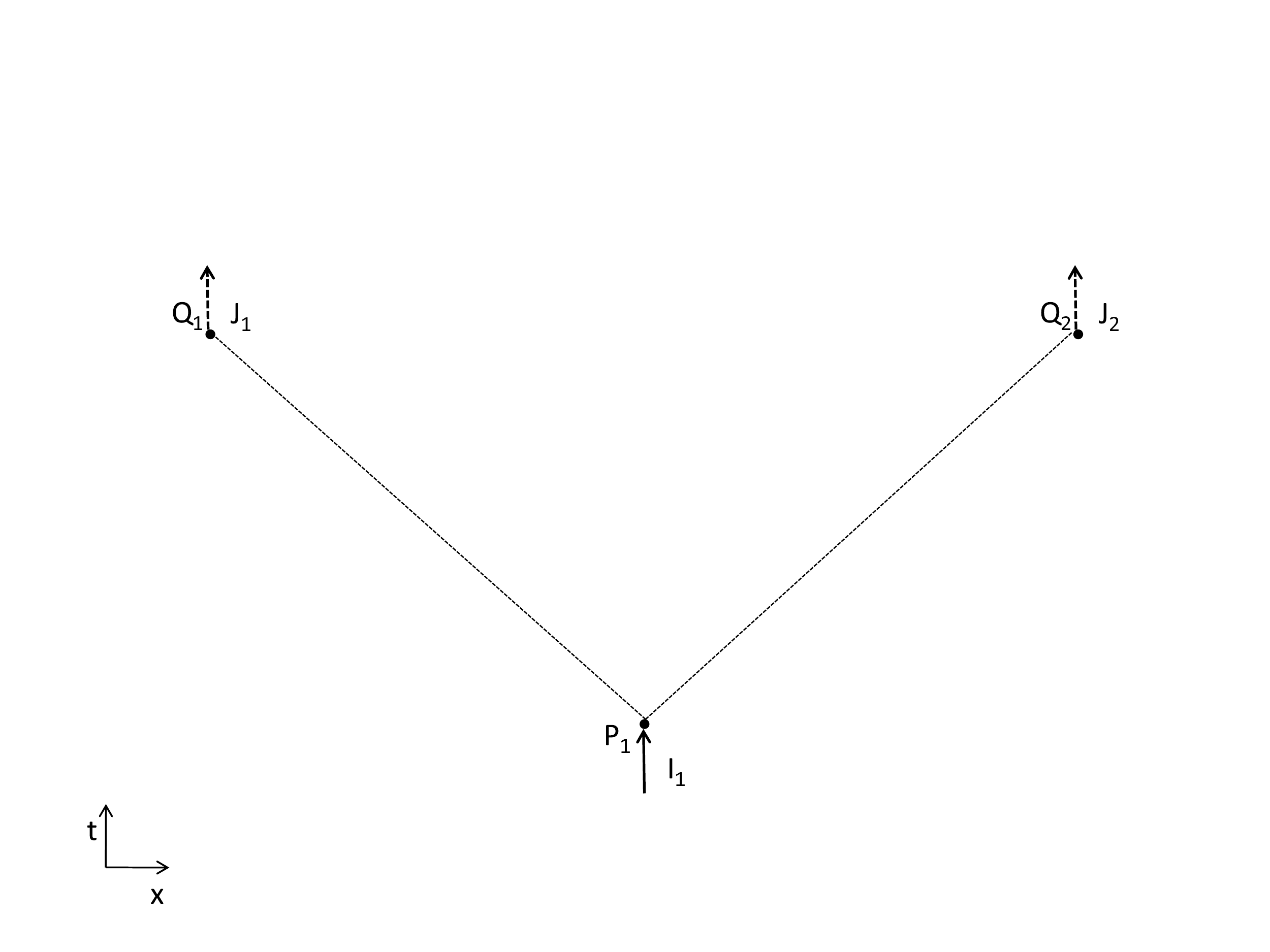}
\caption{Relativistic bit commitment protocols
represented in our framework. 
Alice receives input  $I_1$ at point $P_1$ from Bob.
She is required to output information, $J_1$ and $J_2$,
to Bob's agents at points $Q_1$ and $Q_2$ that are lightlike separated 
from $P_1$ in different directions.  Together, $J_1$ and $J_2$
constitute an unveiling of Alice's commitment, and are intended 
to guarantee to Bob that Alice was indeed committed at the point $P_1$. 
}\end{figure}

\newpage

To explain exactly what these protocols achieve, it is helpful to 
use our framework to model the process by which 
Alice herself learns the bit value $b$
to which she will commit.      
In practice, this bit $b$ might be the result of a computation that 
Alice carries out, or a fact about nature
that she learns, or even a thought (perhaps a prediction based on
intuition) that pops into her mind.   These processes take place
in space and time, and so one might begin by modelling them by
an interaction with an oracle that supplies Alice with the value
$b$ at some definite point $P$ in space-time.   

However, in each
case, the processes can be distributed in space-time.  The bit
can depend on data computed or observed over a large region of space-time,
for example.  Indeed, even an individual's thoughts are in
principle not completely localized, generated as they are by 
neural interactions over a somewhat extended region.  

Moreover, the same computation might be carried out independently
by space-like separated devices.  The same fact
about nature could be inferred by observations
made at many space-like separated points.
And even the same thought 
might occur to several agents independently -- perhaps as a result 
of starting from the same premises, or from observing separate but
correlated events.   It follows that, 
in each case, the bit -- or any partial information about the
bit -- can be generated {\it redundantly}.  

To allow for this, we allow 
Alice to receive inputs from many oracles, from which input data
$b$ can (if necessary by bringing the inputs together at a single
point) be deduced, and we also allow Alice to receive input data
independently many times.  An important special case of this is 
that she may receive the commitment bit $b$ itself independently
many times at space-like separated points.
We call any model that includes input data from which $b$ can 
ultimately be deduced, whether once or redundantly at many 
space-like separated sites, an {\it oracle input model}
for the bit $b$. 

Now suppose that 
an oracle input model $M$ describes Alice's generation of the bit
$b$, and suppose that she unveils $b$ faithfully according to a given
bit commitment protocol,  
We say that the bit commitment protocol guarantees that Alice was
committed {\it by the space-time point} $P$ if any such model $M$ 
necessarily implies that (propagating input information as required)
she could deduce the bit $b$ at the point $P$.  That is, Alice cannot
unveil the bit $b$ {\it unless} it was available to her at $P$.  

To see this is a significant distinction, consider the following
classical bit commitment protocol in $1+1$ dimensions.  (Figure 7.)  Alice is supposed to commit
herself at $P_1$ by sending the bit $b$, encrypted, at light speed
to the points $Q_1$ and $Q_2$.  Her agents at $Q_1$ and $Q_2$ 
unveil the bit $b$ to Bob's agents there by decrypting the signal
and relaying the value of $b$ to them.

\begin{figure}[h]
\centering
\includegraphics[scale=0.4]{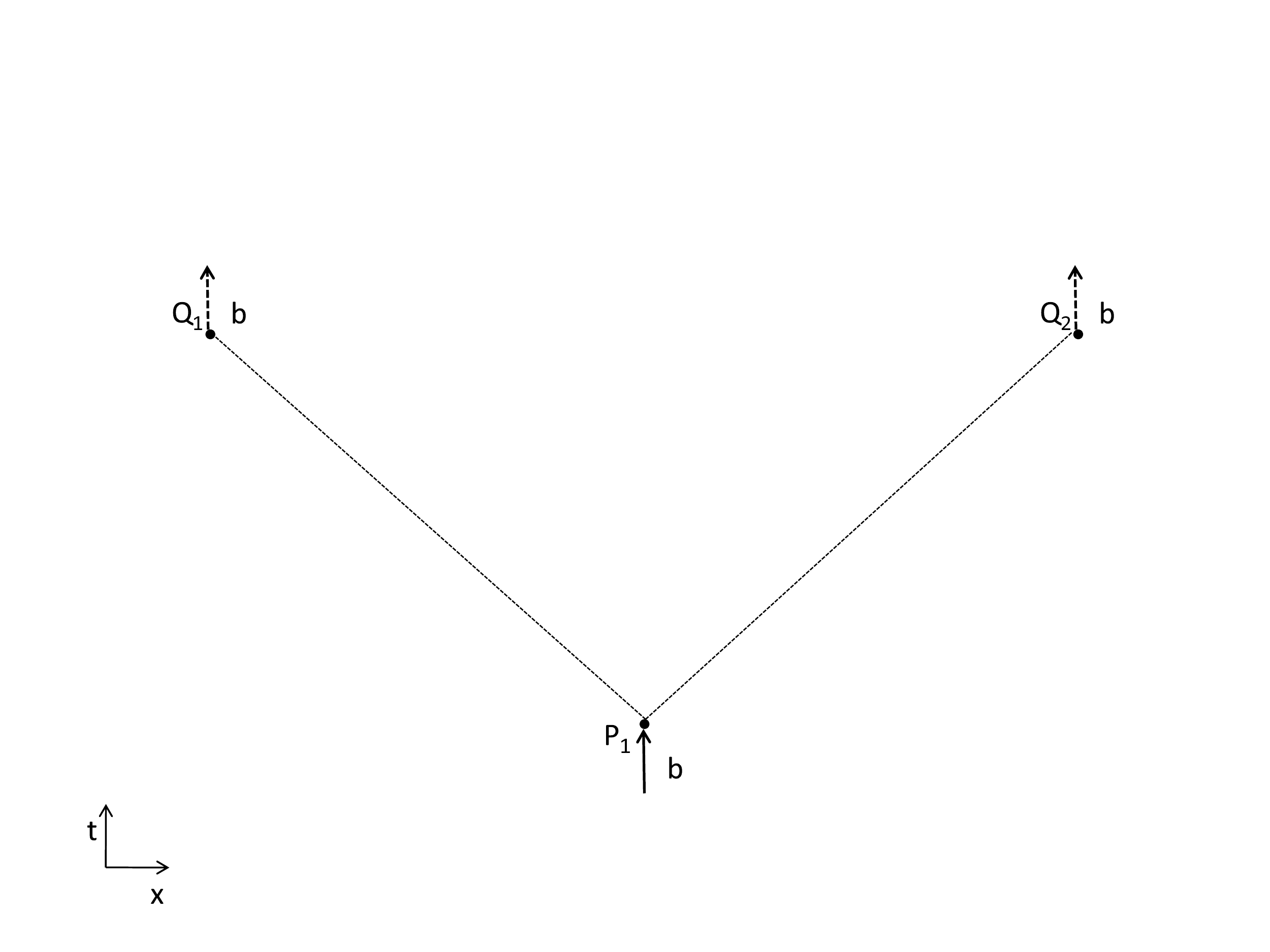}
\caption{A classical relativistic bit commitment protocol in
$1+1$ dimensions 
represented in our framework. 
Alice learns the bit $b$ at point $P_1$.
She is required to send the encrypted bit to her agents 
at $Q_1$ and $Q_2$, points lightlike separated 
from $P_1$ in different directions.  Her agents decrypt the
bit and give it to Bob's agents at $Q_1$ and $Q_2$.  
Note that while this protocol does indeed allow Bob to infer some constraints
on Alice's acquisition of $b$, it does {\it not} guarantee to Bob that she
was committed by the point $P_1$.   
}\end{figure}

\newpage

If Alice's acquisition of the bit $b$ can be modelled by a single
input from a single pointlike oracle at some point $X$, this protocol
guarantees that $X$ must be in the intersection of the past light
cones of $Q_1$ and $Q_2$ and hence (in $1+1$ dimensions) that it must
be in the past light cone of $P_1$.   So, within this restricted model
of bit generation, the protocol indeed would guarantee that Alice must
be committed by the point $P_1$. 

However, our general model allows many other possibilities. 
For example, Alice could receive the bit $b$ independently from
two oracles at points $Q'_1$ and $Q'_2$ that lie on the light
rays $PQ_1$ and $PQ_2$ respectively.  (Figure 8.)  In this case,
she would still be able to comply with the protocol for unveiling
$b$, although the value of $b$ was not known to her at $P_1$. 
In other words, she is not genuinely committed at or before $P_1$. 

This is surely the correct conclusion, by any reasonable definition
of commitment.  For example, Alice could carry out computations of $b$ with
two computers that (in laboratory frame) have the same space
coordinates as $Q'_1$ and $Q'_2$, starting at the time coordinate 
of $P$, so that the computations complete at $Q'_1$ and $Q'_2$ and
supply her agents there with the value of $b$.   Clearly, this does
not imply that she had the bit value (or even the data from which it
is computed) available at $P_1$.

\begin{figure}[h]
\centering
\includegraphics[scale=0.4]{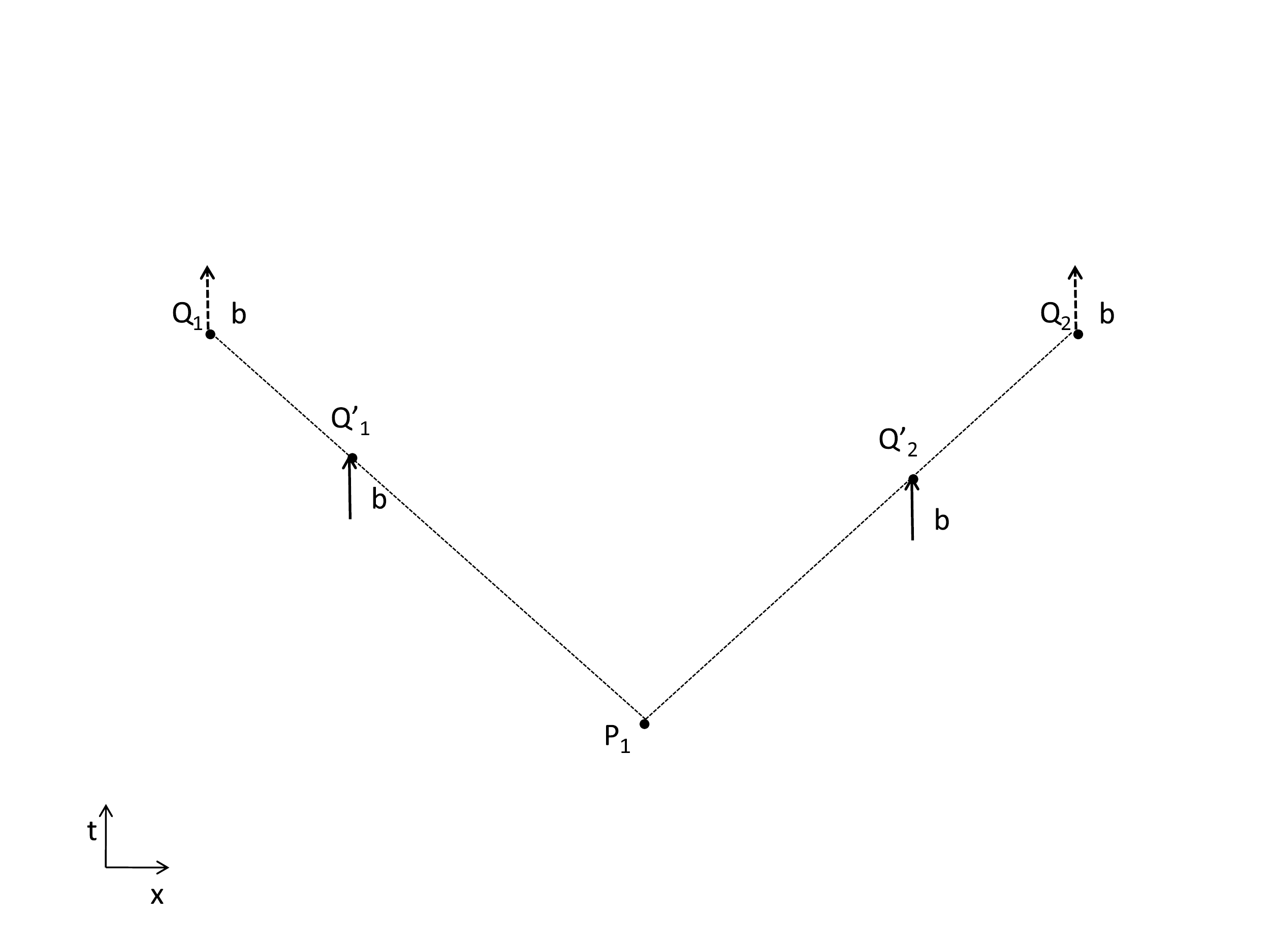}
\caption{Defeating the classical relativistic bit commitment protocol
described in Figure 7.
Alice learns the bit $b$ independently at points $Q'_1$ and $Q'_2$.
She sends the bit to her agents 
at $Q_1$ and $Q_2$, who give it to Bob's agents at $Q_1$ and $Q_2$.  
Alice's unveiling is apparently valid, but she did {\it not} have
the bit $b$ available at the point $P_1$, and so clearly was not
committed there. 
}\end{figure}

\newpage

\subsection{Computational examples}

Quite general questions about distributed computations in Minkowski
space, with classical or quantum inputs and outputs, can be posed
within our framework.   It is easy to construct simple examples of task
which require a non-trivial strategy -- something more than local
computations and direct signalling -- to complete.  

For instance, suppose that Alice is given an unknown qubit as
input at point $P_1$, whose $(x,y,z,t)$ coordinates are $(0,0,0,0)$ in
$3+1$ dimensions, 
where $c=1$.  Suppose that at the 
point $P_2 = (3,0,0,2)$ she is given a second input, in the form
of a classical bit, instructing her to return the qubit either
at the point $Q_0 = (3,4,0,6)$ or to the point $Q_1 = (3,-4,0,6)$.
She cannot complete this task by transmitting the qubit from $P_1$
to $P_2$ and then on to the stipulated $Q_i$, since $P_1$ and $P_2$
are space-like separated.   Nor is there any other
path along which the qubit can be transmitted that guarantees
that she can complete the task. 

Naively, one might take this as 
an argument that it is impossible for Alice to guarantee completing
the task.   However, she {\it can} guarantee to complete the
task, by predistributing an entangled singlet shared between $P_1$
and $P_2$, teleporting the qubit as soon as it arrives at $P_1$,
broadcasting the classical teleportation data in all directions,
transmitting the entangled partner qubit from $P_2$ to the stipulated
$Q_i$, and recombining the classical and quantum teleportation data
at the relevant $Q_i$.   

Teleportation-based attacks  \cite{kms,
buhrmanetal} on quantum tagging schemes
give a large class of similar examples, in which
a party (referred to as Eve in the tagging literature)
can use teleportation to complete tasks that
naively appear impossible (given that, in these cases,
she is excluded from the region occupied by the tag). 

Other examples of non-trivial strategies for completing
relativistic tasks can be constructed by using 
quantum secret sharing \cite{qsecrets} techniques, in which an unknown
state can be effectively non-trivially delocalized and
recombined in a variety of ways.   
These examples suggest that characterizing which distributed computational tasks are possible
and which impossible is not at all a trivial question. 
It seems, on the contrary, almost completely open and very
interesting.  

\section{Excluding Alice from specified regions} 

One cryptographically interesting type of restriction that can be
imposed on Alice in our framework is that she must complete the
task while being excluded from some (not necessarily connected)
region of space-time.    

\begin{figure}[h]
\centering
\includegraphics[scale=0.4]{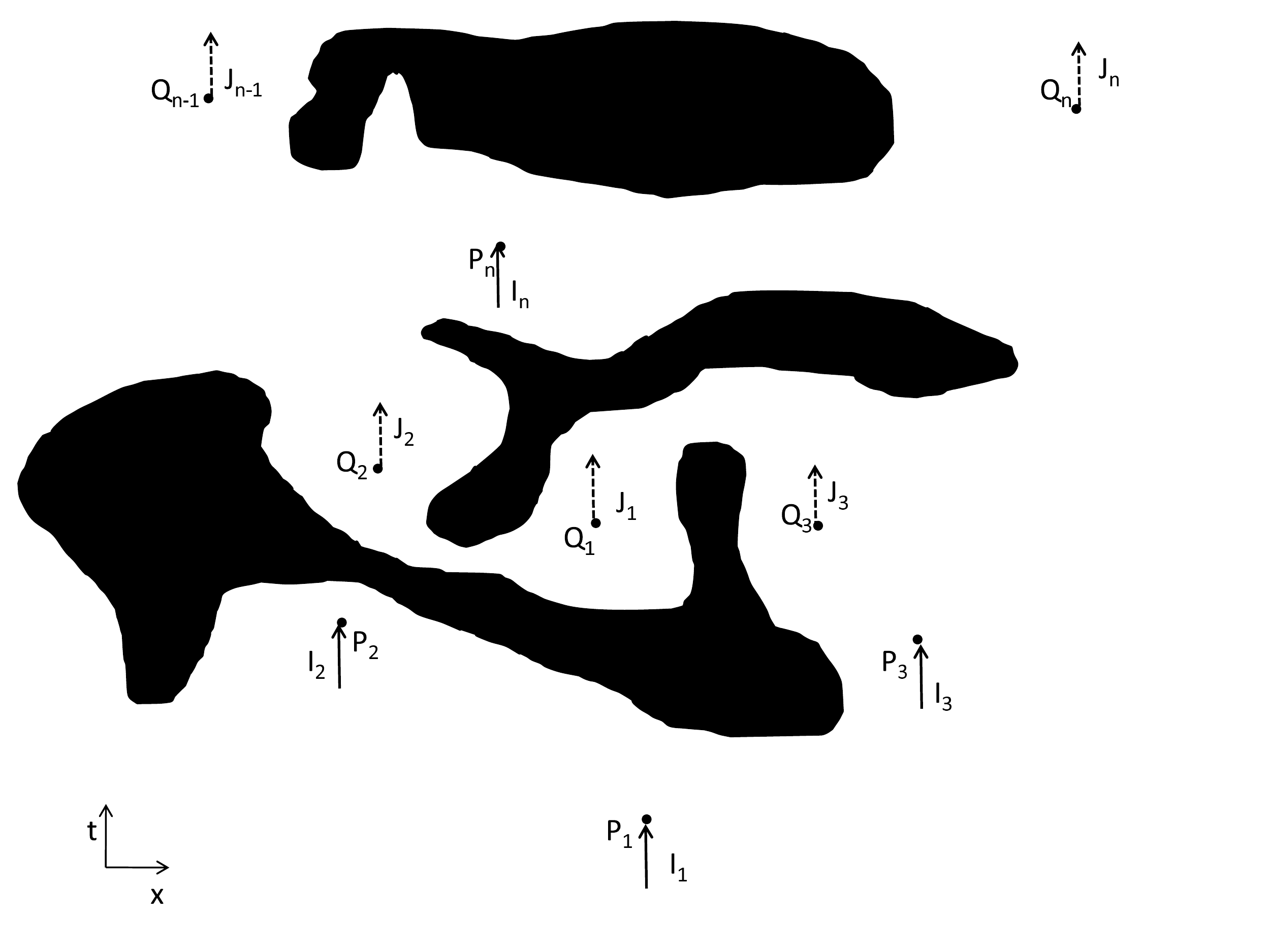}
\caption{An illustration of a relativistic quantum task
in $1+1$ dimensions with restrictions on the location 
of Alice's agents. 
Alice receives inputs $I_1 , \ldots , I_m$ at points $P_1 , \ldots , P_m$. 
Following a prearranged protocol, she is required to 
calculate output points $Q_1 , \ldots , Q_n$ and produce the
output data $J_1 , \ldots , J_n$ there.
Her agents may be located anywhere in space-time except for the
darkened regions.} 
\end{figure}

\newpage

\subsection{Cryptographic applications: quantum tagging in Minkowski space}

In the context of quantum tagging or position authentication, the
underlying idea here is to design tasks that Alice {\it can} complete if she
is allowed agents within a 
particular space-time region but {\it cannot} complete if excluded
from that region.   The completion of the task would then serve as
a guarantee that Alice does indeed have one or more agents located
within the region.   In the simplest and most natural example,
the relevant region is the desired world-tube of some finite object,
the {\it tag}, and the aim is to verify that the tag is indeed
following the desired path.   This is harder to ensure \cite{kms,
buhrmanetal} than one might initially hope \cite{chandranetal,malaney,
kms}.  

Reviewing these intriguing results is beyond our scope here;
interested readers are referred in particular to
Ref. \cite{buhrmanetal} for a strong no-go theorem in one
security model and Ref.
\cite{kenttaggingcrypto} for a strong positive result in another. 

As these references highlight, there is more than one interesting
way of defining relativistic quantum tasks given excluded regions.  
These different definitions point to 
different cryptographically relevant security models, and 
also suggest different ways of probing the properties
of relativistic quantum theory.      

One interesting option is to allow some of the inputs to be at points
{\it within} the proscribed region: cryptographically, this
models a tag that is able to retain and use secret information.
It is intuitively plausible -- and indeed turns out to be
correct \cite{kenttaggingcrypto} -- that this allows us to
define tasks that guarantee secure tagging. 
Alice cannot access these inputs if excluded from
the region, but in general
requires them to generate the required outputs, so
she cannot complete the task in this case.
On the other hand, if she is allowed agents within
the region, she has access to all the necessary inputs,
and so is able to complete sensibly designed tasks.  

A second interesting possibility is to suppose not only that Alice's
agents are excluded from the region, but that her
signals (classical and quantum) also are -- i.e. that the region is effectively
{\it impenetrable}.  While the existence of a region
that is impenetrable to signals may seem a very strong
assumption, it is a standard one in some  cryptographic contexts.
For example, a fully device-independent quantum cryptographic
protocol requires that the devices used in the protocol 
(which are assumed to be
constructed by an adversary, Eve) are contained within secure
laboratories and are unable to send any signal through the 
laboratory walls.  This ensures that the devices cannot 
communicate with Eve and prevents them from being able to
report all their inputs and outputs to her (which would
make any protocol transparent and so make device-independent
cryptography impossible).   In a model in which tags
are able to receive signals on their boundary and propagate
them through their interior if they choose (i.e. if the 
signals are of the right form and arrive at the right point,
according to the tagging protocol), but are otherwise impenetrable,
some useful forms of tagging are possible even with (only) classical
inputs and outputs \cite{akunpublished}. 

\newpage

\subsection{Other mistrustful cryptographic tasks in Minkowski space}

For a cryptographic protocol 
involving two mistrustful parties, Alice and Bob, it is standard to 
assume that each occupies a secure laboratory that they control 
and that the other cannot access or inspect.   The laboratories
are thus disjoint.   For unconditional security, it is also assumed
that the parties trust {\it nothing} outside their own laboratory:
they have to allow for the possibility that everything outside is
under the control of the other party.   

In Minkowski space, it turns out \cite{kentrelfinite} to be valuable to allow each
party's laboratory to be disconnected, so that they control
at least two separated sites.   Alternatively, the laboratory 
can be connected but spatially extended, allowing signals to
be sent and received at two well separated locations within a 
signal laboratory.   In either case, a relativistic cryptographic
protocol for a task involving mistrust will specify (at least
approximately) times and locations at which signals are sent
and received.    

Protocols are, obviously, designed so that 
each party can comply with the protocol: we say a party that
does so is {\it honest}.   A security proof then needs to 
show that (except perhaps with a small probability), the {\it only}
way to appear to comply with the protocol, by producing outputs of
the right form given the inputs received, is to behave honestly.   
For example, a full quantum security proof for the bit commitment protocol
of Ref. \cite{kentrelfinite} would need to show that the committer, Alice,
who is excluded from Bob's two laboratories but potentially in control
of the rest of space-time, can produce outputs of the required form
only by following the protocol for committing a given qubit.
(See Figure 10.)   

\begin{figure}[bh]
\centering
\includegraphics[scale=0.4]{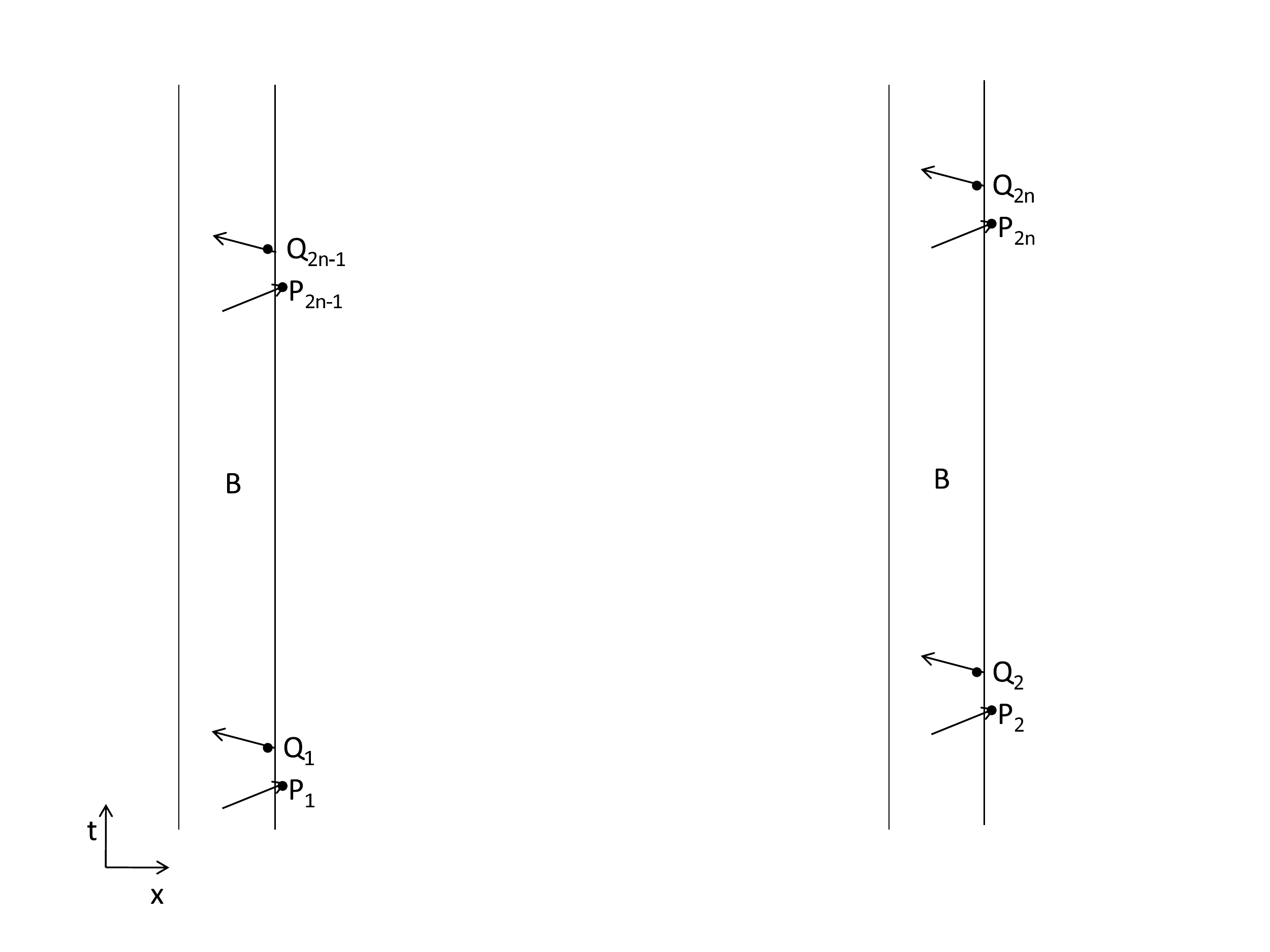}
\caption{The relativistic bit commitment protocol of 
Ref. \cite{kentrelfinite} represented in our framework. 
Alice is excluded from the world-tubes of Bob's secure
laboratories, but is potentially able to site agents
anywhere else in space-time.  Alice receives inputs $I_1 , \ldots ,
I_{2n}$ in the form of queries from Bob, arriving at the points
$P_1 , \ldots , P_{2n}$, where the odd labelled queries come
from one of Bob's laboratories and the even labelled queries from
another.   Each successive pair of points $P_i , P_{i+1}$ is
space-like separated.  To commit to a bit and sustain the
commitment, Alice is required to produce
output data $J_1 , \ldots , J_{2n}$ of a form specified by
the protocol and transmit these data
to arrive at points $Q_1 , \ldots , Q_{2n}$.  
She can complete this task by following the protocol and 
committing to a bit $b$ (or a quantum superposition of bits).
A full security proof for the protocol requires showing that
this is the {\it only} strategy which gives a technologically
unbounded Alice a significant probability of completing the task.}
\end{figure}

\newpage
\section{Other restrictions on Alice's communications}

The principle of information causality \cite{infocaus} suggests
another interesting generalization.   (Figure 11.)  Alice
may be excluded from specified regions, as above, and
her communication through these regions may be constrained,
but not completely excluded.   For example, Alice may be 
restricted to sending a finite number of bits and/or qubits
through any given region.

\begin{figure}[bh]
\centering
\includegraphics[scale=0.4]{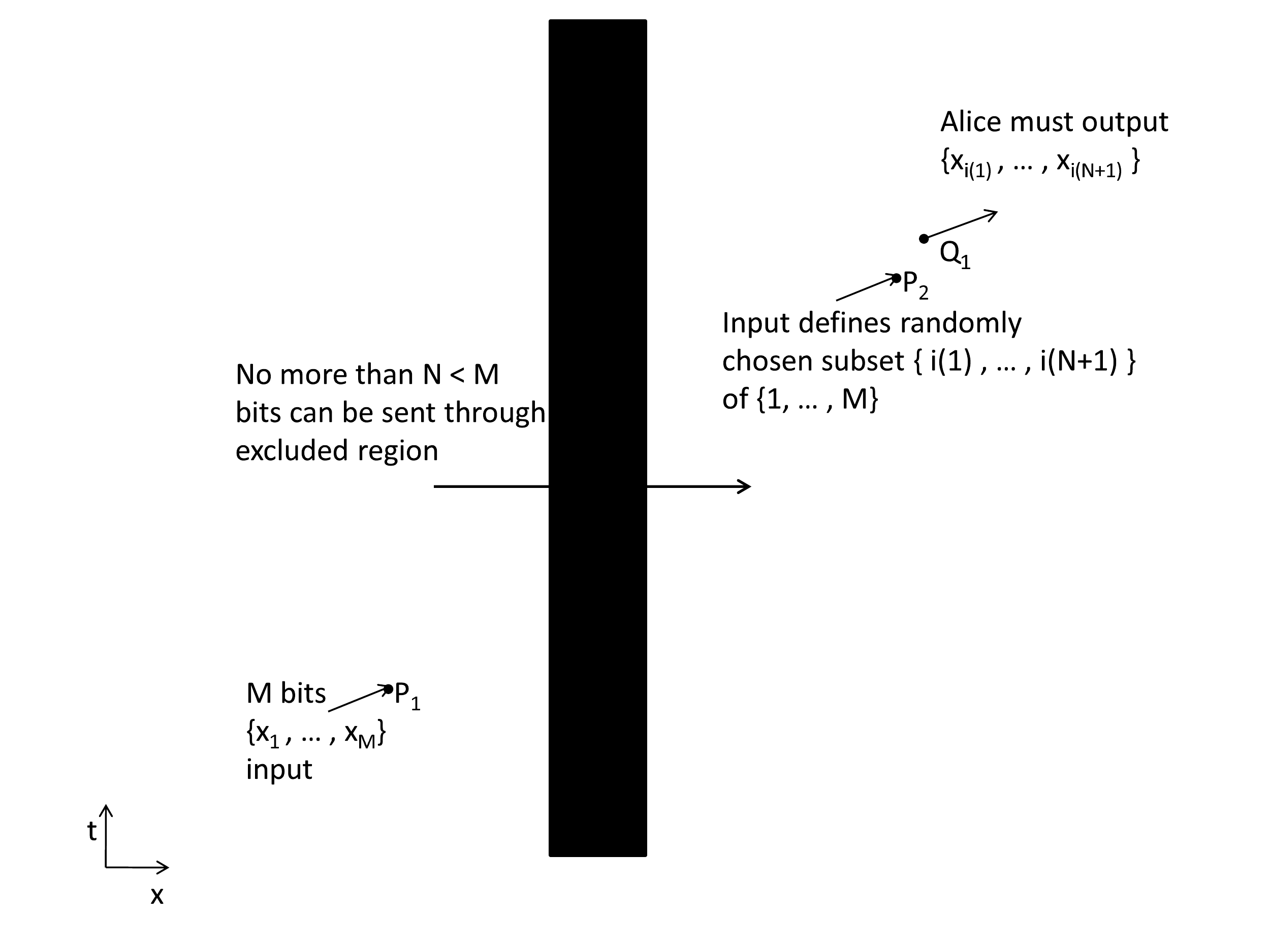}
\caption{The principle of information causality \cite{infocaus}
represented in our framework. 
Alice receives input $I_1$, which takes the form of a string of
$M$ bits, at point $P_1$, and input $I_2$, which takes the form
of a query for $N+1 \leq M$ of the $M$ bits, at point $P_2$. 
She is required to produce the $N+1$ requested bits at
the point $Q_1$.  Her agents may be located anywhere in space-time 
except for the darkened region.  The darkened region is only 
penetrable to a limited extent: she may transmit no more than a total of $N$ 
bits through it.  She cannot generally complete the task.} 
\end{figure}
\newpage

\section{Discussion}

This paper has set out a framework that allows quantum tasks in
Minkowski space to be rigorously defined, and described 
concrete applications to quantum cryptography and computing.
In particular, it not only allows relativistic quantum cryptographic
tasks to be defined rigorously, but also allows a rigorous definition
of the security criteria for these tasks.  

The framework highlights the need for a more systematic understanding of
general principles, such as no-signalling, no-cloning, 
no-summoning and information causality, that allow us to 
characterize which tasks are possible and which impossible.  
We hope it may encourage a wider interest in these intriguing
questions, and more generally in the foundations of relativistic
quantum theory and quantum information.  

In this context, recent work by Coecke \cite{coecke}, 
Hardy \cite{hardytensor} and 
Chiribella et al. \cite{chiribella}  
on abstract frameworks for analysing quantum tasks also deserves
mention.  While these ideas have different motivations and different
mathematical expressions, and address different problems, it would
be intriguing if connections could be drawn.

\acknowledgments
I thank Giulio Chiribella, Bob Coecke, Roger Colbeck, Boris Groisman,
Lucien Hardy, Richard Jozsa, Serge Massar, Graeme Mitchison,
Stefano Pironio, Damian Pitalua-Garcia, Tony
Short and Jonathan Silman for helpful discussions.
This work was partially supported by a Leverhulme Research Fellowship, a grant
from the John Templeton Foundation, and by Perimeter Institute for Theoretical
Physics. Research at Perimeter Institute is supported by the Government of Canada through Industry Canada and
by the Province of Ontario through the Ministry of Research and Innovation.
I also acknowledge the support of the
EU Quantum Computer Science project (contract
255961).


\end{document}